\documentclass[final,5p,twocolumn,times]{elsarticle}

\usepackage{graphicx}
\usepackage{dcolumn}
\usepackage{bm}

\newcommand {\be}{\begin{equation}}
 \newcommand {\ee}{\end{equation}}
 \newcommand {\bea}{\begin{eqnarray}}
 
 \newcommand {\eea}{\end{eqnarray}}

\title{Experimental confirmation of the $\Lambda(1405)$ ansatz
 from resonant formation \\
  of a $K^-p$ quasi-bound state in $K^-$ absorption by $^3$He and $^4$He
  }
\author[rvt,irn]{Jafar Esmaili\corref{cor1}
}
\ead{jesmaili@riken.jp}
\author[rvt,focal]{Yoshinori Akaishi
}
\ead{akaishi@post.kek.jp}
\author[rvt,els]{Toshimitsu Yamazaki
}
\ead{yamazaki@nucl.phys.s.u-tokyo.ac.jp}
\cortext[cor1]{Corresponding author; jesmaili@riken.jp}
\address[rvt]{RIKEN, Nishina Center, Wako, Saitama 351-0198, Japan}
\address[irn]{Department of  Physics, Isfahan University of Technology, Isfahan 84156-83111, Iran}
\address[focal]{College of Science and Technology, Nihon University, Funabashi, Chiba 274-8501, Japan}
\address[els]{Department of Physics, University of Tokyo, Bunkyo-ku, Tokyo 113-0033, Japan}

\begin{document}

\begin{abstract}
 The $\Sigma \pi$ invariant-mass spectra in the resonant capture of $K^-$  at rest in $^4$He, $^3$He and $d$ are calculated by a coupled-channel procedure for a $K^-p$ quasi-bound state of an arbitrary chosen mass ($M$) and width ($\Gamma$). A $\chi^2$ analysis of old $^4$He bubble chamber data shows a dominance of the s-orbit absorption, and yielded
$M = 1405.5^{+1.4}_{-1}$ MeV/$c^2$ and $\Gamma = 26 ^{+4}_{-3}$ MeV, where a possible population of $\Sigma^0(1385)$ and also a small p-orbit capture contribution are taken into account. This result confirms the $\Lambda(1405)$ ansatz, whereas recent chiral-SU(3) predictions ($M \sim 1420$ MeV/$c^2$) are excluded. A more stringent test by using a $^3$He target is proposed .
\end{abstract}


\maketitle

\noindent
\section{Introduction}

A quest for antikaon-nuclear bound states has become a
hot topic in nuclear hadronic physics. In recent years we have predicted deeply bound
kaonic states, and studied their structure and formation based on the $\bar{K}N$
interaction, which was derived empirically by a coupled-channel treatment of the $\bar{K}N$ and $\pi \Sigma$ channels so as to account for
the known low-energy $\bar{K}N$ quantities \cite{AY02PRC,AY02PLB,AY04PLBa,AY04PRC,AY04PLBb}.
This Akaishi-Yamazaki (AY) interaction is based on a traditional ansatz that the
$\Lambda(1405)$ resonance is the bound $I=0$ $\bar{K} N$ state. The strong binding regime as a natural consequence of the $\Lambda(1405)$ ansatz leads to the prediction of a deeply bound $K^-pp$ and others, yielding a super-strong nuclear force \cite{Yamazaki:07a,Yamazaki:07b}, and eventually kaon condensed matter. Experimental evidence for a deeply bound $K^-pp$ system in favour of a strong binding regime
has been obtained recently \cite{Yamazaki:08}.

Theoretically, M\"uller-Groeling, Holinde and Speth \cite{Speth:90} predicted a strongly attractive $I=0 ~\bar{K} N$ interaction from a meson-exchange treatment, and Waas, Kaiser and Weise \cite{Waas:96} derived a similarly strong $\bar{K} N$ interaction from a chiral SU(3) model, which is consistent with the $\Lambda(1405)$ ansatz, thus leading to a strong binding scheme (we call this model ``Chiral-Strong"). On the other hand, in recent years, some
theories starting from the chiral SU(3) dynamics, but with a different approach (zero-range and strong energy dependence) lead to a much less-attractive $\bar{K} N$ interaction, claiming
that the $K^-p$ state is located at $Mc^2 \sim 1421 - 1434$ MeV \cite{ORB,HNJH,BNW,BMN,JOORM,HW} together with a second pole which is mainly coupled with $\Sigma \pi$.  Hereafter, such a  ``theoretical" state will be called symbolically ``$\Lambda^*(1420)$". This new theoretical consequence, which we call ``Chiral-Weak",  leads to a weak binding regime of kaonic nuclear states, and no deeply  bound kaonic nuclear states are expected.
 Although a theoretical account against the ``Chiral-Weak" regime will be given elsewhere \cite{Akaishi:09}, it is urgently important to ask a crucial question: where is the $K^- p$ resonance state - $\Lambda(1405)$ or $\Lambda^*(1420)$? In the present Letter we propose experimental methods to distinguish these cases from resonant formation of $\Lambda(1405)$ or $\Lambda^*(1420)$ in $K^-$ capture at rest by $^3$He and $^3$He and show some  evidence for $\Lambda(1405)$ from $K^-$ absorption by $^4$He.

The present-day PDG value of $\Lambda (1405)$ \cite{PDG} depends heavily on theoretical arguments presented by Dalitz and Deloff (hereafter called DD91) \cite{Dalitz91}. They chose exclusively 10 data points below the $\bar KN$ threshold among Hemingway's $\Sigma^+\pi^-$ invariant mass spectrum \cite{Hemingway85}, and searched for the $\chi^2$ minimum in $|T_{\Sigma \pi,\Sigma \pi}|^2$ fitting as a function of the resonance energy, $E_{\rm R}$, under a constraint of an $I=0~\bar KN$ scattering length. The obtained resonance-pole location is distributed as $1405-i27$ MeV ($M$ matrix), $1387-i40$ MeV ($K$ matrix) and $1526-i159$ MeV (SR potential). DD91  expressed a strong preference for the $M$-matrix model, and recommended a value of $(1406.5 \pm 4.0)-i(25 \pm 1)$ MeV, which is taken up as the PDG value. We strongly think that an "entrance-channel ambiguity", that is, ``$|T_{\Sigma \pi,\Sigma \pi}|^2$ fit or $|T_{\Sigma \pi,\bar KN}|^2$ fit", must be settled for an analysis of Hemingway's data, since the $|T_{\Sigma \pi,\Sigma \pi}|^2$ fit deviates seriously from the data above the $\bar KN$ threshold \cite{Akaishi:09}.

We now point out that the invariant-mass spectrum of $\Sigma^{\pm} \pi^{\mp}$ from stopped $K^-$ in $^4$He by Riley {\it et al.} \cite{Riley} is equally, or even more, valuable for determining the $\Lambda (1405)$ resonance position. So far, since the data shape looks like a quasi-free (QF) spectrum, it has not been used for the purpose of deducing information about $\Lambda(1405)$. In this Letter, however, we clarify that the spectrum comes essentially from the resonant formation of $\Lambda (1405)$, but is  a "projected invariant-mass spectrum" governed by the momentum distribution of the spectator, $^3$H, of the $K^-p \rightarrow \Sigma \pi$ process. We search for the $\chi^2$ minimum for $|T_{\Sigma \pi, K^-p}|^2$ as a function of the pole energy, $M_{\rm pole}$, and width, $\Gamma$. A great advantage of our analysis is that it has no complicated problem concerning the "entrance-channel ambiguity". Another merit is that the property of the $\Lambda^*$ resonance in the $\bar KN$ channel is well reflected in the  $|T_{\Sigma \pi, K^-p}|^2$ spectrum, in contrast to the $|T_{\Sigma \pi,\Sigma \pi}|^2$ one, where serious interference takes place between the continuum and the resonance.

The essence of the present Letter is to formulate $K^-$ absorption by $^4$He, $^3$He and $d$ as a resonant capture of $K^-$ by a {\it nuclear proton} ($``p"$) to form an assumed $K^- p$ state of any given mass, $M_{\rm pole}$, and to compare the calculated $\Sigma \pi$ invariant-mass ($M_{\Sigma \pi}$) spectra with experiments. For this purpose,
we treat the $K^-$ capture as direct and resonant capture processes:
\bea
K^- + ``p" &\rightarrow& \Sigma + \pi {\rm ~~(direct~capture,~``DC")},\\
                \rightarrow \Lambda^* &\rightarrow&  \Sigma + \pi {\rm ~~(resonant~capture,~``RC")},
\eea
where ``$p$" is a bound proton in a target nucleus with a binding energy value of $B_p$. We can tune the
$\Lambda^*$ resonance by using nuclei with different $B_p$ values. The non-resonant direct capture process
(called QF) can also contribute to $M_{\Sigma\pi}$. We investigate this problem in detail.\\


\section{$K^-p$ state as a Feshbach resonance}

For a model setting we treat the $K^-p$ quasi-bound state as a Feshbach resonance \cite{Feshbach58}, which is embedded in the continuum of $\Sigma \pi$. As given previously \cite{Akaishi:08}, we consider two channels of $\bar KN$ and $\pi\Sigma$ for simplicity. We  employ a set of separable potentials with a Yukawa-type form factor \cite{Yamaguchi54},
\begin{eqnarray}
\langle \vec k' \mid v_{ij} \mid \vec k \rangle = g(\vec k') ~U_{ij} g(\vec k), ~~~g(\vec k) = \frac {\Lambda^2} {\Lambda^2 + \vec k^2}, \\
U_{ij} = \frac{1}{\pi^2} \frac{\hbar^2}{2 \sqrt{\mu_i \mu_j}}
\frac{1}{\Lambda} s_{ij}, \hspace{2cm}
\label{YYint}
\end{eqnarray}
where $i(j)$ stands for the $\bar KN$ channel, 1, or the $\pi \Sigma$ channel, 2, $\mu_i(\mu_j)$ is the reduced mass of channel $i(j)$, and $s_{ij}$ are non-dimensional strength parameters. Then, a complex potential with the following strength is derived analytically:
\begin{eqnarray}
s_{1}^{\rm {opt}}(E) &=& s_{11}-s_{12} \frac {\Lambda^2}{(\Lambda-i\kappa_2)^2+s_{22}\Lambda^2} s_{21}, \nonumber\\
\frac {\hbar^2}{2 \mu_2} \kappa_2^2 &=& E+\Delta Mc^2,
\label{sopt}
\end{eqnarray}
where $\Delta M = m_{K^-} + M_p - m_{\pi^{\pm}} - M_{\Sigma^{\mp}} =99$ MeV$/c^2$ is the threshold mass difference, and  $\kappa_2$ is a complex momentum in the $\pi \Sigma$ channel. The complex energy, $E_{\rm{pol}}$, of the  {\it pole} state for the three interaction parameters ($s_{11}$, $s_{12}$ and $s_{22}$) is obtained by solving
\begin{equation}
E_{\rm{pol}} = \Xi (E_{\rm{pol}}),
\end{equation}
where
\begin{equation}
\Xi (z) \equiv - \frac {\hbar^2}{2 \mu_1} \Lambda^2 (\sqrt{- s_1^{\rm {opt}}(z)} - 1)^2.
\label{E0}
\end{equation}
Conversely, for a given value of $E_{\rm pole}$, two of the strength parameters can be calculated. The choice of $s_{22}$ is arbitrary in this model, as long as the binding energy and the width are concerned (this trivial point is often misunderstood as if the AY treatment \cite{AY02PRC} were invalid \cite{HW}).
We adopt $s_{22} = -0.66$, which gives $U_{22}/U_{11}= 4/3$ for $\Lambda(1405)$ as in a ``chiral" model, and $\Lambda$ = 3.90 /fm.  For example, the two assumptions for the $K^-p$ state are found to correspond to the following parameters:
\begin{eqnarray}
\Lambda(1405)&&s_{11}=-1.28,~s_{12}=0.28,~s_{22}=-0.66,\\
\Lambda^*(1420)&& s_{11}=-1.17,~s_{12}=0.32,~s_{22}=-0.66.
\end{eqnarray}
In the following treatment we obtain $s_{11}$ and $s_{12}$ from the $M$ and $\Gamma$ values of an arbitrary chosen $K^-p$ state to be used to calculate the $\Sigma \pi$ invariant masses.

It should be remarked that the invariant mass that we treat here is never the genuine invariant mass of the parent $X$, since $X = K^-p$ is produced in a kinematically constrained way \cite{AY99PLB}. It is a ``partial" invariant mass, as in the Dalitz presentation of three final particles (1, 2 and 3) from a parent. In the present case of $K^-$ absorption at rest,
\bea
K^- + ^4{\rm He} &\rightarrow& \Sigma + \pi + t,\\
K^- + ^3{\rm He} &\rightarrow& \Sigma + \pi + d,\\
K^- + d &\rightarrow& \Sigma + \pi + n,
\eea
the invariant-mass ($M_{\Sigma\pi}$) distribution is very much constrained by the kinematics, and can be called a ``projected" invariant-mass distribution. \\

\begin{figure*}[htb]
\begin{center}
 \includegraphics[width=17cm]{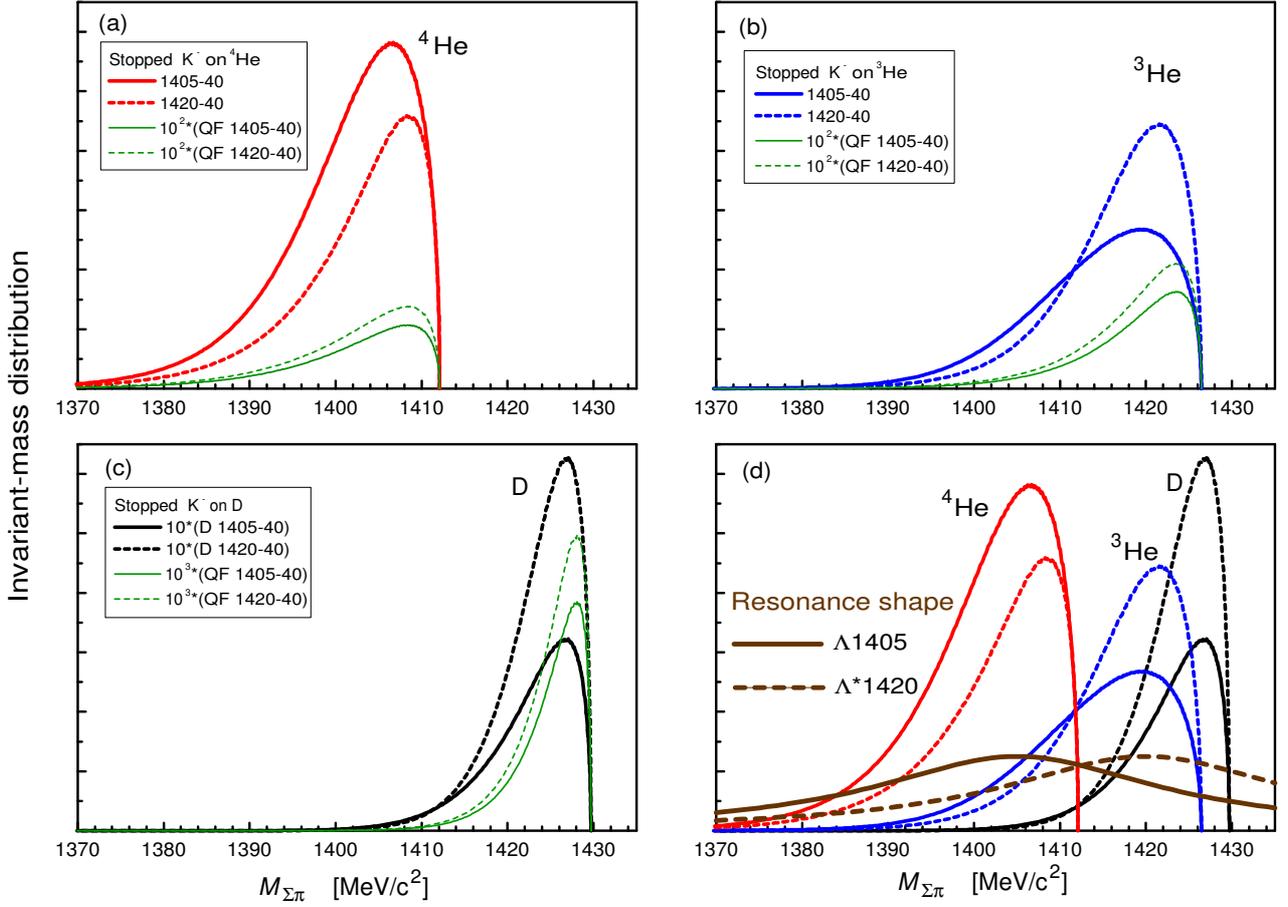}\\
  \caption{Resonant and QF $M_{\Sigma \pi}$ spectra in $K^-$ absorption from the s-orbit by
  (a) $^4$He (red), (b) $^3$He (blue) and (c) $d$ (black) for $M_{\Lambda^*}$ = 1405 (solid curve) and 1420 (broken curve)
  with HO potential. They are compared
  with the resonance shapes (free invariant masses) for $\Lambda(1405)$ and $\Lambda^*(1420)$ in (d).}\label{fig:He34d}
\end{center}
\end{figure*}


\section{$\Sigma\pi$ invariant-mass from stopped $K^-$ on He nuclei}

\subsection{Formulation}

The coupled-channel scattering amplitude for
the elementary process, $T_{ij}$, satisfies
\be T_{ij}=U_{ij}+ \sum_l \, U_{il} \, G_{l} \,T_{lj}\ee
with Green's function, $G_{l}$. The solution is given in a matrix form by
\begin{center}
$T=[1-UG]^{-1}U$.
\end{center}
In our treatment Green's function is considered to be
\be (UG)_{ij}=-s_{ij}\sqrt{\frac{\mu_j}{\mu_i}}\frac{\Lambda^2}{(\Lambda-ik_j)^2},\ee
where $k_j$ is a relative momentum in the channel $j$.

In our calculation, we treat a single proton bound in a target nucleus with a binding energy $B_p$ and the remaining part of the nucleus as a spectator  ($S$), following \cite{Yamazaki:07}. We have considered a potential between them so that it reproduces the experimental binding energy for the $p+S$ system.
The momentum distribution of the decay particles in the $K^-$ s-orbit absorption is given as
\bea
&& \frac{d^2\Gamma}{dk_\Sigma dk_S} =\frac{2(2\pi)^3}{\hbar^2c^2} \mid \psi_{nlm}^{\rm{atom}}(0)
 \mid^2 \nonumber\\
&& \times \mid g(k^{'})T_{21}(E_2)g(\frac{1}{2}k_S) \mid^2 k_\Sigma k_S E_\pi \mid F(k_S)\mid^2, \label{eq:momentum-distribution}\eea
 \be E_2=\sqrt{(E_i-E_S)^2-\hbar^2c^2k_S^2}-M_{\Sigma}c^2-m_{\pi}c^2,\ee
where $\psi_{nlm}^{\rm{atom}}(0)$ is a $K^-$ atomic wave function; $k_\Sigma$, $k_\pi$ and $k_S$  are the momenta of
$\Sigma$, $\pi$ and the spectator $S$, respectively, and the $T_{21}$ involves
the $\Lambda^*$ resonance effect. The kinematical constraints among the various momenta are given by
\be k^{'}=\sqrt{k_{\Sigma}^2+\frac{1}{4}k_S^2+k_{\Sigma}k_Sx}, \ee
\be k_{\pi}=\sqrt{k_{\Sigma}^2+k_S^2+2k_{\Sigma}k_S x}, \ee
\be x=\frac{(E_i-E_{\Sigma}-E_S)^2-(m_{\pi}^2c^4+\hbar^2 c^2 (k_{\Sigma}^2 + k_S^2))}{2 \hbar^2 c^2 k_\Sigma k_S},  \ee
where $x=\cos\theta_{\Sigma S}$ and $|x|\leq 1$ is kinematically allowed.
In our treatment, the quasi-free spectrum is produced by $U_{21}$ in the place of $T_{21}$.

The invariant mass can be reconstructed from the momenta of the daughter particles. In the present case, since the daughters are only three particles ($\Sigma$, $\pi$ and $S$), $M_{\Sigma\pi}$ is identical to the spectator missing mass, $\Delta M(S)$. The spectator momentum distribution is calculated as
\be \frac{d\Gamma}{dk_S}=\int_0 ^\infty dk_\Sigma \frac{d^2 \Gamma}{dk_\Sigma dk_S}.\ee
The $\Sigma \pi$  invariant mass
distribution is given by
\be \frac{d\Gamma}{d(M_{\Sigma\pi}c^2)}=\frac{E_S}{\hbar^2c^2k_S}\frac{\sqrt{E_i^2+M_S^2c^4-2E_iE_S}}{E_i}\frac{d\Gamma}{dk_S}, \ee
where $M_S$, $k_S$ and $E_S$ are the mass, momentum and energy of the spectator, respectively,  and $E_i$ is the initial energy of the kaonic atom.\\

\begin{figure}[htb]
\begin{center}
   \includegraphics[width=8cm]{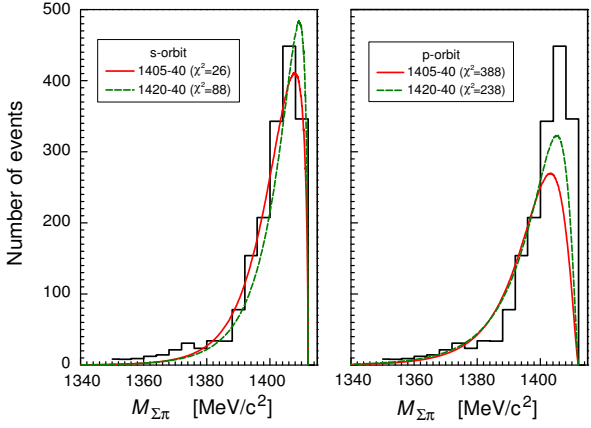}\\
  \caption{Comparison of a $\Sigma^{\mp}\pi^{\pm}$ invariant-mass spectrum of Riley {\it et al} \cite{Riley} from $K^-$ stopped on $^4$He with best-fit theoretical curves of s and p-orbit absorption  with
   the Harada potential for $\Lambda(1405)$ and $\Lambda^*(1420)$ and $\Gamma$ = 40 MeV. }\label{fig:s-p-orbits}
\end{center}
\end{figure}

\subsection{$M_{\Sigma\pi}$ distributions}

We calculated the $M_{\Sigma\pi}$ spectra from $K^-$ absorption in $^4$He, $^3$He and $d$ for resonant capture from  the s-orbit by $\Lambda(1405)$ and $\Lambda^*(1420)$ as well as for direct capture (QF; non-resonant). Figure~\ref{fig:He34d} overviews the characteristic properties of the $M_{\Sigma\pi}$ distribution. Here,
 the harmonic oscillator (HO) potential is used as the
 interaction potential between the proton and the spectator in each target.

\begin{figure}[htb]
\begin{center}
 \includegraphics[width=8cm]{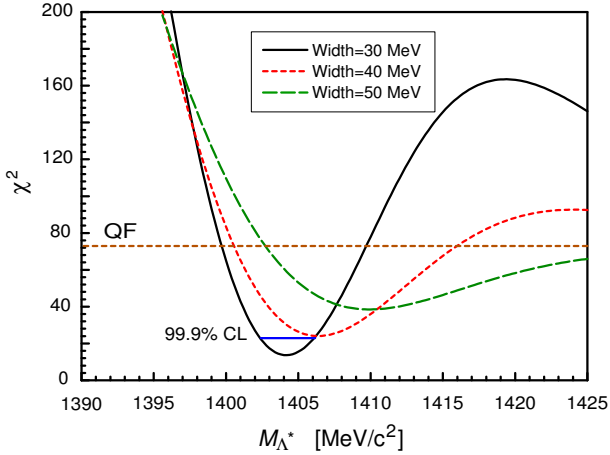}\\
  \caption{$\chi^2$ distributions versus $M$ from the $\Sigma^{\mp}\pi^{\pm}$ invariant-mass spectrum of stopped $K^-$ on $^4$He \cite{Riley} best fitted
  to theoretical curves with the s-orbit absorption mode adopting the Harada potential for $\Gamma$ = 30, 40, 50 MeV.}\label{fig:chisq-M}
\end{center}
\end{figure}

The resonance shapes of $\Lambda(1405)$ and $\Lambda^*(1420)$ are shown in Fig.~\ref{fig:He34d}d) by bold brown solid and broken curves, respectively.
The $M_{\Sigma\pi}$ is peaked just below the kinematical limit given by each target. This results from a small  momentum distribution of the spectator. The spectrum (\ref{eq:momentum-distribution}) is governed and projected by the spectator momentum distribution, $|F(k_S)|^2$, because it is sharper than the resonance shape, as reflected in $T_{21} (E_2)$. This property makes all of the spectra similar to each other, no matter where the resonant capture occurs. In the case of $^4$He, the QF spectrum and the resonant-capture spectrum look very much similar in shape, and both are close to the observed spectrum, which reveals a peak with its position at around 1405 MeV. At first glance, it might indicate $\Lambda(1405)$ formation, but it can also be interpreted as a QF spectrum. Thus, we need a careful look into the problem to solve the issue of $\Lambda(1405)$ versus $\Lambda^*(1420)$.

Indeed, the energy of $K^- + ^4$He at rest (4221 MeV) nearly coincides with the energy of $\Lambda(1405) + ^3$H (4216 MeV), so that the resonant-capture condition is well fulfilled. The theoretical spectrum is shown in Fig.~\ref{fig:He34d}a) (red curve). On the other hand, the energy of $\Lambda^*(1420) + ^3$H (4229 MeV) is about 15 MeV higher (``off tuned"), but its corresponding spectrum (red broken curve) has a similar shape, while its intensity is reduced. Both ``resonant-capture" spectra are similar, and the QF  spectra, which have much smaller intensities, also have similar shapes. This situation might indicate that it is difficult to solve the issue of $\Lambda(1405)$ versus $\Lambda^*(1420)$ from $M_{\Sigma\pi}$, but we will show in the following that a good experimental spectrum is capable of distinguishing these small differences.

It is interesting to point out that the resonance condition with $\Lambda^*(1420)$ is fulfilled in the $^3$He target, since the energy of $K^- + ^3$He at rest (3302 MeV) is close to that of $\Lambda^*(1420) + ^2$H (3296 MeV), whereas the energy of $\Lambda(1405) + ^2$H (3283 MeV) is off. As shown in Fig.~\ref{fig:He34d}b), the spectrum with $\Lambda^*(1420)$ (blue broken curve) is larger in intensity than that with $\Lambda(1405)$ (blue solid curve), and the latter is characterized by a long tail toward the lower mass, where the resonant-capture is enhanced. We propose a future experiment to detect this significant difference in shape between $^3$He and $^4$He to distinguish between $\Lambda(1405)$ and $\Lambda^*(1420)$.

\subsection{The s-orbit absorption favoured}

Hereafter we analyze the bubble-chamber data of stopped-$K^-$ absorption in $^4$He by Riley {\it et al.} \cite{Riley}. We use a
sum of the two spectra, $N_i^{\Sigma^- \pi^+}$ and $N_i^{\Sigma^+ \pi^-}$, consisting of $n = 9$ data points in the range of 1378 to 1410 MeV/$c^2$. Generally, the experimental histogram $N_i, i = 1,,,n$ with statistical errors $\sigma_i = \sqrt{N_i}$ is fitted to a theoretical curve, $S(x; M_{\rm pole}, \Gamma)$ with $x = M_{\Sigma \pi}$ involving the mass $M_{\rm pole}$ and width $\Gamma$ as two parameters, by minimizing the $\chi^2$'s value:
\begin{equation}
\chi^2 (M_{\rm pole}, \Gamma) = \sum^n_{i=1} \left( \frac{N_i - S(x_i; M_{\rm pole}, \Gamma)}{\sigma_i} \right)^2
\end{equation}

First, we examine from which atomic orbit the $K^-$ nuclear capture takes place.
The calculated $M_{\Sigma\pi}$ spectra in $^4$He for the s-orbit and the p-orbit absorption are fitted to experimental data, as shown in Fig.~\ref{fig:s-p-orbits}. We observe that the best-fit $\chi^2$ value for the s-orbit capture ($\chi^2 = 26$ for $M_{\rm pole} = 1405$ MeV/$c^2$and $\Gamma =40$ MeV) is not far from the expected $\chi^2 \sim  \, n_{\rm DF} \pm \sqrt{2 \, n_{\rm DF}} = n-1 \pm \sqrt{2 \, (n-1)} = 12 \pm 5$. On the other hand, the best-fit $\chi^2$ value is much larger for the p-orbit capture ($\chi^2 = 388$), and thus, the data favours the s-orbit capture. This is consistent with the known fact that negative mesons and antiprotons in liquid helium, when captured in large-$n$ atomic orbits, undergo s-orbit capture after Stark mixing decays (around 97\% \cite{pbar}). $K^-$ mesons are captured by atomic states with high principal quantum numbers (23$\sim$28), from where
the $K^-$  proceeds to lower states, and ultimately reaches s-states of various principal quantum numbers.\\

\subsection{Precise comparison in the case of $^4$He}

 Now let us examine more quantitatively whether or not we can really distinguish between the $\Lambda(1405)$ and $\Lambda^*(1420)$ capture processes, and furthermore attempt to determine $M_{\rm pole}$ of the $K^- p$ state.
In this work we have used another potential for describing the interaction between $p$ and $t$  in $^4$He.
It is a Gaussian $3N-N$ potential, $U_{N}(\vec{R})$, which was derived
from a microscopic four-body calculation by Harada \cite{Harada}. The Harada potential is parameterized into
useful Gaussian forms, and can reproduce the experimental data of the binding energy, $B_{N}=20.6$ MeV, for the $t+p$ system.

So far, we have not taken into account the $I=1$ component of the $\bar{K} N$ interaction. We now evaluate its effect by using R$\acute{\rm e}$vai and Shevchenko's interaction \cite{Revai09}, which accounts for the $K^-p \rightarrow \Sigma^{\pm} \pi^{\mp} $ available data: it is known that their ratio reflects the interference of the $I=0$ and $I=1$ amplitudes \cite{Staronski87}. We find that the contribution of the $I=1$ component to the whole spectral intensity is about 4\%. This affects the results of our $\chi^2$ fitting only slightly: the $M_{\rm pole}$ value changes by 0.2 MeV, and the $\Gamma$ value by 0.4 MeV. So, we can safely neglect the $I=1$ effect within the deviation.

\begin{figure}[htb]
\begin{center}
\includegraphics[width=8cm]{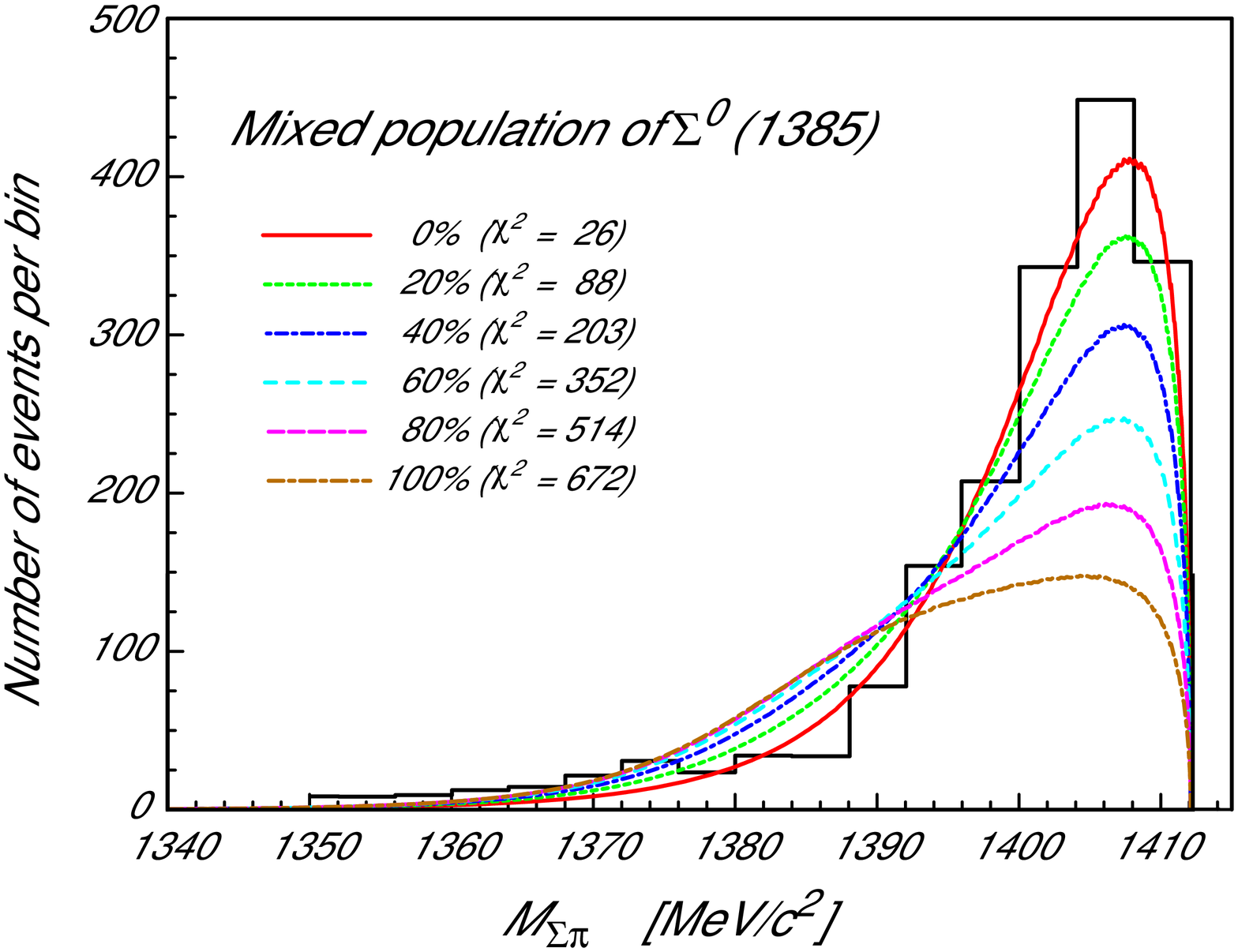}\\
  \caption{(Theoretical $M_{\Sigma^{\mp}\pi^{\pm}}$ spectra of mixed $\Sigma^0(1385)$ population fitted to the experimental spectrum. $\chi^2$ values are given.}\label{fig:L*S*-mixed}\label{fig:L*S*-mixed}
\end{center}
\end{figure}

We performed $\chi^2$-fitting to the $K^{-4}$He data. The $\chi^2$ values were obtained from the best-fit procedures of the experimental data to theoretical $M_{\Sigma\pi}$ spectra with an assumed resonance mass ($M_{\rm pole}$) and a given width ($\Gamma$ = 30, 40 and 50 MeV), as shown in Fig.~\ref{fig:chisq-M}. A very significant minimum, $\chi^2 _0 \equiv \chi^2 (M_{\rm pole})$, is observed at around $M_{\rm pole} = 1405$ MeV/c$^2$ in the $\chi^2$ distribution. The excess $\chi^2 (x)$ distribution, $\Delta \chi^2 (x) = \chi^2 (x)- \chi^2 (x_0)$, is generally given by a Poisson distribution $P_1 (x) = x \, {\rm exp} (-x)$, and thus, the increment ($x$) of $\chi^2$ distributes as
\begin{equation}
L(x) = 1 - (1+x) \, {\rm exp} \,(-x).
\end{equation}
The increments, $\Delta \chi^2 (x) =$ 2.36, 3.89, 4.74, 6.64 and 9.23, correspond to confidence levels of 68.3 (one standard deviation), 90, 95, 99 and 99.9 \%, respectively. The case of $\Lambda^*(1420)$ is excluded.

We made a more general fitting including $\Gamma$ as an additional parameter, and obtained a contour of $\chi^2$ in the plane of $M_{\rm pole}$ and $\Gamma$, as shown in Fig.~\ref{fig:confidence-contour}. The $\chi^2$ minimum and its 1$\sigma$ (68\% confidence level) give $M_{\rm pole}$ = 1404.9 $^{+2.4}_{-1.4}$ MeV/c$^2$ and $\Gamma$ = 35.5 $^{+7.6}_{-5.4}$ MeV for the s-orbit absorption case.\\

\subsection{Effects of the $\Sigma(1385)$ resonance and the p-orbit capture}

We examined the effect of the population of the $\Sigma^0(1385)$ resonance, which is known to decay to $\Sigma \pi$ with a branching of about 12\% \cite{PDG}. Figure \ref{fig:L*S*-mixed} shows $\chi^2$ fitting results with assumed populations of 0 to 100\%.  The best-fit value of the mixing is 0 - 20\%. The mixing effect is to shift the centroid of the contour curves toward a larger $M$ value. Thus, to be safe, we adopt 10\% mixing.

\begin{figure}
\begin{center}
 \includegraphics[width=9cm]{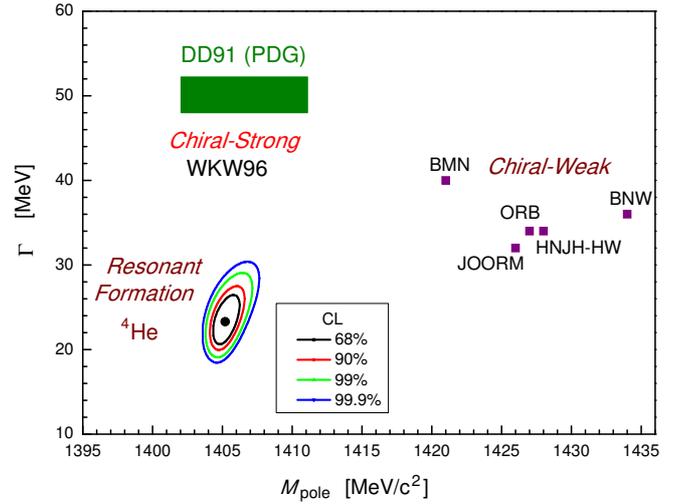}
  \caption{$M-\Gamma$ contour curves of the confidence levels for the $\chi^2$ fitting of $M_{\Sigma^{\mp}\pi^{\pm}}$ spectrum from $K^-$ stopped on $^4$He. The effects of the $\Sigma^0(1385)$ population (10\%) and of the p-orbit capture (10\%) are taken into account. The predicted values of ``Chiral-Weak" models together with the DD91 (PDG) and the ``Chiral-Strong" zone are shown.}\label{fig:confidence-contour}
\end{center}
\end{figure}

A small mixing of the p-orbit capture was also taken into account, because this will make a shift of  $M$ toward a higher value. Finally, we made a contour plot with a 10\% mixing of the p-orbit capture and a 10\% mixing of the $\Sigma^0(1385)$ population. This is already taken into account in Fig.~\ref{fig:confidence-contour}.

\begin{figure}
\begin{center}
 \includegraphics[width=9cm]{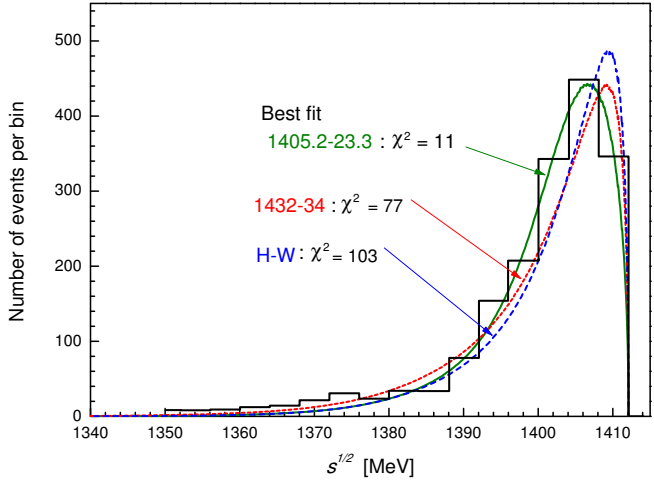}
  \caption{Detailed differences in $M_{\Sigma \pi}$ spectra among the Hyodo-Weise prediction and the present model predictions. } \label{fig:HW}
\end{center}
\end{figure}

\section
{Conclusion \label{conclusion}}

The $M-\Gamma$ contour presentation of the confidence level of our fitting of the old invariant-mass spectrum of $\Sigma^{\mp}\pi^{\pm}$ from $K^-$ absorption in $^4$He \cite{Riley} by our theoretical curves shows a deep minimum ($\chi^2 \sim 11$), giving
\be
M = 1405.5^{+1.4}_{-1.0}~{\rm MeV}/c^2~ {\rm and}~ \Gamma = 25.6 ^{+4}_{-3}~{\rm MeV}.
\ee
This fitting procedure takes into account the effects of p-orbit mixing to 10\% and of the $\Sigma^0(1385)$ population to 10\%. The above $M$ value is in good agreement with the known ones \cite{PDG} and is consistent with the ``Chiral-Strong" prediction. It contradicts seriously the $\Lambda^*(1420)$ ansatz. The predicted $M$ and $\Gamma$ values of the ``Chiral-Weak" models are located far outside the 99.9\% confidence contour, and thus, are definitely incompatible with the experiment.

We have used in the above $\chi ^2$ analysis the separable potential model of Eq.(2.1-5), not with the chiral models.  In order to estimate the difference between two types of models, we also calculated $\chi^2$ fitting for Hyodo-Weise's two-channel model of the chiral SU(3) dynamics. As shown in Fig.~\ref{fig:HW}, the obtained $\chi^2$ value ($\sim$103) is much larger than the best fit case ($\chi^2=11.4$). It is even larger than $\chi^2 \sim 77$ at a corresponding pole position by our procedure on the map, $M_{\Lambda^*}=1432$ MeV and $\Gamma = 34$ MeV. This unfavorable feature comes from the strong energy-dependence of the Weinberg-Tomozawa term which makes the spectrum dropping more rapidly toward the lower invariant-mass region (see Fig. \ref{fig:HW}). This is a general tendency of ``Chiral-Weak" models. Thus, we can safely conclude that ``Chiral-Weak" models are located far outside the 99.9\% confidence level.

The width that we have obtained seems to be significantly smaller than the usually believed ones. We emphasize that the PDG value based on DD91 has little confidence. There are no other experimental data to deduce the width more reliably than the present one.
We have shown that the $K^-$ capture by $^3$He provides even a better testing ground, because this nucleus offers better tuning of resonant capture to $\Lambda^*(1420)$, if this exists at all. On the other hand, the $M_{\Sigma \pi}$ spectrum will show a long tail extending to the $\Lambda(1405)$ region, if it exists. Such an experiment of $K^-$
absorption at rest on $^3$He, which can be done at J-PARC and DA$\Phi$NE, is highly awaited.

The case of a $d$ target, as shown in Fig.~\ref{fig:He34d}c), indicates that $\Lambda^*(1420)$ has an effect near the threshold. We notice that the $\Lambda(1405)$ resonance, which is far from the threshold, can be populated as a nearly isolated peak when we consider the realistic momentum distribution of $d$. This will be explained in a forthcoming paper \cite{Esmaili-KD}.

\section{Acknowledgements}

The authors are grateful to Professor P. Kienle for his stimulating discussion. One of us (J.E.) would like to thank Professor E. Hiyama for inviting him to RIKEN and the hospitality during the time this work was performed, and thank Professor S. Z. Kalantari for helping him before coming to Japan. This work is supported by the Grant-in-Aid for Scientific Research of Monbu-kagaku-sho of Japan.

\end{document}